\documentclass[aps,twocolumn,amssymb,showpacs]{revtex4}

\usepackage{graphicx}% Include figure files

\newcommand{\be}{\begin{equation}}
\newcommand{\ee}{\end{equation}}
\newcommand{\bea}{\begin{eqnarray}}
\newcommand{\eea}{\end{eqnarray}}

\newcommand{\f}{\psi}
\def\k{\kappa}
\def\r{\rho}
\newcommand{\half}{\frac{1}{2}}

\newcommand{\p}{\partial}

\begin{document}

\title{Backreaction in Acoustic Black Holes}
\author{Roberto Balbinot, Serena Fagnocchi }
\altaffiliation{Email addresses: balbinot@bo.infn.it, fagnocchi@bo.infn.it}
\affiliation{Dipartimento di Fisica dell'Universit\`a di Bologna
and INFN sezione di Bologna, \\  Via Irnerio 46,
40126 Bologna, Italy }
\author{Alessandro Fabbri}
\altaffiliation{Email address: fabbria@bo.infn.it}
\affiliation{Departamento de Fisica Teorica, Facultad de Fisica, Universidad de 
Valencia, \\ Burjassot-46100, Valencia, Spain}
\author{Giovanni P. Procopio}
\altaffiliation{Email address: gpp27@cam.ac.uk}
\affiliation{DAMTP, Centre for Mathematical Sciences, University of Cambridge
\\ Wilberforce road, Cambridge CB3 OWA, UK}

\begin{abstract}
The backreaction equations for the linearized quantum fluctuations 
in an acoustic black hole are given. The solution near the horizon, obtained 
within a dimensional reduction, indicates that acoustic black holes, unlike
Schwarzschild ones, get cooler as they radiate phonons. They show remarkable
analogies with near-extremal Reissner-Nordstr\"om black holes.
\end{abstract}


\maketitle

One of the most surprising and far reaching result for its implications in modern theoretical physics is 
the prediction made by Hawking \cite{hawking} that black holes emit thermal radiation at a temperature 
$T_H$ proportional to the surface gravity $k$ of the horizon. For a Schwarzschild black hole of mass 
$M$, $k=(4M)^{-1}$ and $T_H=\hbar (8\pi M)^{-1}$ (we have set the velocity of light and Boltzman constant equal to one).
Hawking obtained this result using quantum field theory in curved space, a scheme for dealing with the matter-gravity system where matter is quantized according to quantum field theory whereas gravity is treated classically according to Einstein General Relativity. The scale at which this framework becomes unreliable is the Planck length where the description of spacetime as a continuous differentiable manifold probably breaks down.
Coming back to black holes, because of the quantum emission, they are unstable. Extrapolating Hawking's result (which is strictly valid only for stationary or static black holes) one can conjecture that as the mass decreases, the hole gets hotter and hotter (being the temperature inversely proportional to the mass) and eventually disappears in a time scale of the order of the initial mass to the third power.
A more quantitative analysis can be performed by looking at the first order (in $\hbar$) corrections 
$g_{\alpha\beta}^{(1)}$ to a classical black hole metric $g_{\alpha\beta}^{(0)}$ induced by the quantum emission. These can be calculated using the semiclassical Einstein equations \cite{York}
\be\label{Einsteinsemiclassiche}
G_{\mu\nu}(g_{\alpha\beta}^{(0)}+g_{\alpha\beta}^{(1)})=8\pi\langle T_{\mu\nu}(g_{\alpha\beta}^{(0)})\rangle.
\ee
Here $G_{\mu\nu}$ is the Einstein tensor evaluated for the quantum corrected metric 
$g_{\alpha\beta}=g_{\alpha\beta}^{(0)}+g_{\alpha\beta}^{(1)}$ 
and linearized in the perturbation $g_{\alpha\beta}^{(1)}$  
(of order $\hbar$). The r.h.s. represents the expectation value of the stress tensor for the quantum matter field 
evaluated in the classical background $g_{\alpha\beta}^{(0)}$.\\ 
In  a very interesting paper, appeared in 1981, Unruh \cite{unruh81} showed that a 
thermal radiation similar to the one predicted by Hawking for black holes is 
expected in a completely (at first sight) different physical scenario, namely a 
fluid undergoing hypersonic motion. This opened the way for the study of condensed 
matter analogues of Hawking radiation \cite{libro}, a rather promising field of 
research where the connections to the experimental side do not seem so remote, 
compared to gravity.

The Eulerian equations of motion for an irrotational and homentropic fluid flow can be 
derived from the action \cite{schakel}, \cite{stone}
\be
\label{Sclassic}
S=-\int d^4 x \left[ \r \dot{\f}+\half \r (\nabla \f)^2+u(\r)\right] ,
\ee
where $\r$ is the mass density, $\f$ the velocity potential, i.e. 
$\overrightarrow{v}=\overrightarrow{\nabla}\f$,  $u$ the internal energy density 
and a dot means time derivative. 

Varying the action with respect to $\f$ one gets the continuity equation
\be
\label{A}
\dot{\r}+\overrightarrow{\nabla}\cdot (\r \overrightarrow{v})=0\ ,
\ee
whereas variation with respect to $\r$ gives Bernoulli's equation ($\mu=\frac{du}{d\r}$)
\be
\label{B}
\dot{\f}+\frac{\overrightarrow{v}^2}{2}+\mu(\r)=0\ .
\ee
\\ Expanding the fields around a solution ($\rho_0$, $\vec{v}_0$) of the classical equations of motion 
(\ref{A},\ref{B}), Unruh showed that the quadratic action $S_{2}$ for the fluctuations of
the velocity potential can be written in a simple and elegant geometrical form, 
namely
\be
\label{sgr}
S_2=-\half \int d^4 x \sqrt{g^{(0)}} g^{(0)\mu\nu} \p_\mu \f_1 \p_\nu \f_1\ ,
\ee
where $\f_1$ is the fluctuation and $g_{\mu\nu}^{(0)}$ is the so called acoustic metric
\be
\label{acousticmetric}
g_{\mu\nu}^{(0)}=-\frac{\r_0}{c}\left(
\begin{array}{cc}
c^2-v^2_0&\overrightarrow{v_0}^T\\
\overrightarrow{v_0}&-I
\end{array}
\right)
\ee
expressed in terms of the background quantities.
$c$ is the sound speed, i.e. 
$c^2=\r d\mu/d\r$, and $I$ is the three-dimensional identity matrix.\\ 
As it can be seen, $S_{2}$ has exacly the same form of an action for a 
massless, 
minimally coupled, scalar field propagating in a ``curved spacetime" 
whose line 
element is $ds^2=g_{\mu\nu}^{(0)}dx^\mu dx^\nu$. 
The region of the fluid for which $v^2_0>c^2$ is called sonic black hole: 
its boundary $|\overrightarrow{v_0}|=c$ defines the sonic horizon. 
Sound waves cannot excape from this region, since they are dragged by the fluid. 
\\A typical example is the Laval nozzle of Fig.1. The fluid flows from right to 
left. At the waist of the nozzle the fluid velocity reaches the speed of sound: 
this is the location of the sonic horizon for free fluid motion. 

\begin{figure}
\includegraphics[angle=270,width=2.5in,clip]{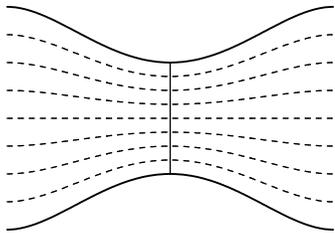}
\caption{A Laval nozzle. The waist of the
nozzle represents the sonic horizon ($|\vec v_0|=c$). In the region on the right of the waist $|\vec v_0|<c$
and on the left $|\vec v_0|>c$ (sonic hole).}
\label{fig1}
\end{figure}

%\begin{figure}
%\centerline{\psfig{figure=laval.ps,width=2.6in,height=1.7in,angle=-90}}
%\end{figure}
%\begin{center}
%\makebox[8.5cm]{\parbox{8.5cm} {\small FIG.1. A Laval nozzle. The waist of the
%nozzle represents the sonic horizon ($|\vec v|=c$). In the region on the right of the waist $|\vec v|<c$
%and on the left $|\vec v|>c$ (sonic hole). }}
% \end{center}

Using Hawking's arguments Unruh, quantizing the field $\f_1$, showed that
in the formation of a sonic hole
one expects a thermal emission of phonons at a temperature $T_U=\hbar k/(2\pi c)$, where $k$ is the surface gravity of the sonic horizon \cite{visser}
\be
\label{k}
k=\left. \half \frac{d}{dn}(c^2-\overrightarrow{v_0}^2)\right|_H \ ,
\ee
$n$ is the normal to the horizon.

In this paper we will give a first qualitative analysis of the effects this 
emitted radiation has on the fluid dynamics, 
i.e., the backreaction of the linearized phonons in sonic black  holes. 
\\ Using standard background field formalism we write the fundamental quantum fields $\hat{\rho}$, $\hat{\f}$ as the sum 
of background fields $\rho$, $\f$ (not necessarily satifying the classical equations of motion), 
plus quantum fluctuations, 
i.e. $\hat{\rho}=\rho+\delta\hat{\rho}$, $\hat{\f}=\f+\delta\hat{\f}$. Integrating out the quantum 
fluctuations one  obtains the 
one loop effective action formally defined as $\Gamma=S+\half \hbar \ \mbox{tr} \ln [\Box_{ g(\rho,\vec v)}]+ O(\hbar^2)$, 
where $S$ is the action (\ref{Sclassic}) and $\Box_{ g(\rho,\vec v)}$ is the covariant D'~Alambertian calculated from an acoustic 
metric $g_{\mu\nu}(\rho,\vec v)$ of the functional form of eq.(\ref{acousticmetric}) 
with $\r_0$ and $v_0$ replaced by $\r$  
and $v$ respectively \cite{bavili}. 
We assume that divergences in the determinant of the above effective action are removed employing a covariant regularization scheme. 
This is the key hypothesis of our work. 
We will comment on it later. Therefore one can write $\Gamma=S+S_q(g_{\mu\nu}(\r,v))$, where the quantum part of the action 
$\Gamma$ depends on the dynamical variables $\r$, $v$ via the acoustic metric only, and coincides with the effective action \
for a massless scalar field propagating in a spacetime whose metric is $g_{\mu\nu}(\r,v)$. 
Using the chain rule we write $\frac{\delta S_q}{\delta \r}=\frac{\delta S_q}{\delta g^{\mu\nu}}\frac{\delta g^{\mu\nu}}{\delta \r}$
and similarly for $\frac{\delta S_q}{\delta \vec v}$.
We write $\r=\r_o+\r_q$ and $v=v_o+ v_q$, where $\r_o,v_o$ satisfy the classical equations of motion 
(\ref{A},\ref{B}) and $\r_q,v_q$ are the corrections of order $O(\hbar)$ induced 
by the quadratic fluctuations of the phonons field. 
The linearized backreaction equations, analogues of semiclassical Einstein eq.(\ref{Einsteinsemiclassiche}), 
read (assuming for simplicity sake a constant velocity of sound $c$)
\bea
\label{eqbackA}
&\dot{\r_q}+\nabla_i(\r_q v_{0i})+\nabla_i \left(\r_0 (v_{qi}-\frac{\langle T_{ti}\rangle}{c^2}
-\frac{v_{0j}}{c^2}\langle T_{ji}\rangle )
\right)  =0\  \ \ \  &\\
\label{eqbackB}
&\dot{\f_q}+\vec{v_0}\cdot \vec{v_q}+\frac{c^2}{\r_0}\r_q- \half \frac{\r_0}{c}\langle T\rangle=0\ , &
\eea 
where $\langle T_{\mu\nu}\rangle \equiv -\frac{2}{\sqrt{-g}}\left.\frac{\delta 
S_q}{\delta g^{\mu\nu}}\right|_{g^{(0)}_{\mu\nu}}$ is the expectation value of the quantum version 
of the so called ``pseudo energy momentum tensor" 
$T_{\mu\nu}=-\frac{2}{\sqrt{-g}}\frac{\delta S_2}{g^{\mu\nu}}$ \cite{stone} and
$\langle T\rangle = g^{(0)\mu\nu}\langle T_{\mu\nu}\rangle$. 
%Note that all these expectation values are evaluated on $g^{(0)}_{\mu\nu}(\r_0,v_0)$, being $\r_0,v_0$ solution of the classical equations of motion.\\
From the above construction it follows that $\langle T_{\mu\nu}\rangle$ coincides with the
expectation values of the stress tensor for a massless scalar field propagating in an effective
spacetime whose metric is $g_{\mu\nu}^{(0)}(\rho_0, v_0)$. 
These expectation values should be taken in the sonic analogue of the Unruh state \cite{unruh76} 
(the quantum state appropriate to describe black hole evaporation at late times), in which, in the 
remote past prior to the time dependent formation of the sonic hole, the quantum field is in its 
vacuum state.
%The adopted choice of covariant regularization implies $\langle T^{\mu\nu}_{\ \ \ ;\nu}\rangle =0$. 
%This equation is the quantum counterpart of the ``covariant conservation law" $T^{\mu\nu}_{\ \ \ ;\nu}=0$ obeyed by the classical 
%pseudo energy momentum tensor. 
%In fact it has been shown (see the second of \cite{stone})  that various known laws of classical fluid motion can 
%be rewritten in an elegant geometrical way as  $T^{\mu\nu}_{\ \ \ ;\nu}=0$. 
%Furthermore these conservation laws may all be interpreted in terms of 
%semiclassical motion of phonons. This gives some support to the regularization scheme we adopted.\\ 
Inspection of the backreaction equations reveals that eq.(\ref{eqbackA}) is the first order 
in $\hbar$ conservation equation for the Noether 
current associated to the $\f\rightarrow \f +const$ symmetry of the effective action $\Gamma$. Note that the phonons contribution 
in the Noether flux coming from $S_q(g_{\mu\nu})$ is just the pseudo momentum density 
$\sqrt{g^{(0)}}\langle T^t_i \rangle$.
% as in the two 
%fluid hydrodynamics of superfluids of Landau and Khalatnikov \cite{volovik}. 
Eq.(\ref{eqbackB}) is the first order Bernoulli equation modified by the presence of the trace term 
which represents an additional contribution to the chemical potential induced by the fluctuations which, 
as we shall see, causes the fluid to slow down. 
These equations reflect the underlying two-component structure of the system as in the
Landau-Khalatnikov theory of superfluidity \cite{khalatnikov}.  
%One can compare the central role played by the trace in the quantum dynamics
%of the fluid with the gravitational black hole case (eq.(\ref{Einsteinsemiclassiche}) where the basic backreaction 
%equation which determines the evolution of the horizon relates the mass loss to the energy flux 
%(in spherical symmetry $\dot M \propto
%\langle T^r_{\ t}\rangle$). In the latter case Hawking radiation occurs at the expense of the gravitational energy of the black
%hole, whereas in the fluid phonons emission takes away kinetic energy from the system.  
One can also rewrite the equation (\ref{eqbackA}) as the usual fluid continuity equation by a simple
redefinition of the velocity field $v_{qi}\rightarrow v_{qi}-\left( \frac{\langle T_{ti}\rangle}{c}+\frac{v_{0j}}{c}
\langle T_{ji}\rangle\right)$ which incorporates the phonon momentum density. 
Consequently the Bernoulli equation (\ref{eqbackB}) rewritten in terms of this redefined velocity contains, besides the trace, 
also other terms.\\
Unfortunately no explicit solutions of eqs.(\ref{eqbackA},\ref{eqbackB}) can be given 
since $\langle T_{\mu\nu} \rangle$ is unknown.\\
In the black hole case, where similar difficulties arise,  
a qualitative insight in the evaporation process can be gained using 
2D dimensional models \cite{strominger}. 
The most popular is the one proposed by Callan, Giddings, Harvey and Strominger (CGHS) \cite{cghs} 
in which the 4D quantum stress tensor is replaced by a 2D one associated to a minimally coupled
massless scalar field described at the quantum level by the Polyakov action \cite{bd}.
%Furthermore transverse pressure terms are neglected.   
%One considers only one spatial dimension, 
%replaces the 4D stress tensor by a 2D one, which in principle can be computed explicitly 
%(or suitably approximated), and neglects transverse pressure terms. This approximation is known to be quite reasonable near the black hole horizon.
\\ With the same spirit a qualitative description of the backreaction 
in a hypersonic fluid can be obtained assuming a one dimensional flow for 
the fluid, let's say along the axis of the nozzle, the $z$ direction. 
So all physical fields will depend only on $t$ and $z$.\\
The effective action for the CGHS-like model for the fluid quantum dynamics can be then given as 
\be
\Gamma^{(2)}=S^{(2)}+ S_{pol}
\ee
where 
\be
S^{(2)}=-\int d^2 x A\left[ \rho\dot\psi + \frac{1}{2}\rho (\partial_z\psi)^2 + u(\rho)\right] \ee
is obtained integrating $S$ over the transverse coordinates $x, y$ and $A$ is the area of the transverse
section of the nozzle.  $S_{pol}$ is the Polyakov action. The backreaction equations following from
$\Gamma^{(2)}$ are 
% The backreaction equations for the dimensionally reduced model, neglecting the transverse pressure, are
\bea
&A\dot{\r}_q+\partial_z [A(\r_q v_0+\r_0 v_q)]& \nonumber \\  
\label{back1}
&-\frac{\partial_z}{c} \left[ (\langle T^{(2)}_{tz}\rangle 
+v_0\langle T^{(2)}_{zz}\rangle )\right]=0& \\
\label{back2}
&A\left( \dot{\psi}_q+v_0 v_q+\frac{c^2}{\r_0}\r_q \right)
- \left[ \frac{\langle T^{(2)}\rangle }{2}\right]=0.&\ \ \ 
\eea
% $A$ is the area of the transverse section. 
% $\langle T_{ab}^{(2)} \rangle $ is the pseudo stress tensor for the dimensionally reduced quantum theory described by the fluctuations action 
% \be
% S_2^{(2)}=-\half \int d^2 x A \frac{\r}{c}\sqrt{-g^{(2)}} g^{(2)ab}\p_a\f_1\p_b\f_1\ ,
% \ee
% which is obtained by integrating (\ref{sgr}) over the transverse directions. Finally $g_{ab}^{(2)}$ is the ($t,z$)-section of the acoustic metric $g_{\mu\nu}^{(0)}$ of eq.(\ref{acousticmetric}). According to our assumption of covariant regularization, being $S_2^{(2)}$ Weyl invariant, $\langle T^{(2)}\rangle$ is exactly known, given by the trace anomaly \cite{grav}
% \be
% \label{traccia}
% T^{(2)}=\frac{\hbar}{24 \pi}\left[ R^{(2)}-6(\nabla\phi)^2+6\Box\phi\right]\ ,
% \ee
% where $R^{(2)}$ is the Ricci scalar of the metric $g_{ab}^{(2)}$ and the ``dilaton" $\phi$ is defined as $e^{-2\phi}=\frac{A\r}{c}$.\\
% The other components of $\langle T_{ab}\rangle$ are not known exactly. 
% As for the black hole case they will be approximated by the corresponding values calculated 
% in the Polyakov theory \cite{bd}. This slightly overestimates the phonon contribution in 
% eq.(\ref{back1}) by neglecting the scattering due to the 
% potential barrier induced by the dilaton (grey-body factor) \cite{bafa5}.\\
$\langle T_{ab}^{(2)}\rangle$ is the quantum stress tensor for a massless scalar field minimally
coupled to the ($t,z$) section $g_{ab}^{(2)}$ of the acoustic metric $g_{\mu\nu}^{(0)}$
of eq. (\ref{acousticmetric}). 
The trace $\langle T^{(2)}\rangle$ is completely anomalous and given by 
\be \langle T^{(2)}\rangle = \frac{\hbar}{24\pi}R^{(2)} \ , \ee where $R^{(2)}$ 
is the Ricci scalar for the metric $g_{ab}^{(2)}$.  
The phonons expectation values appearing in the conservation equation (\ref{back1}) 
can be easily expressed by a coordinate
transformation in terms of $\langle T_{\pm\pm}\rangle$ where $x^{\pm}=t\pm z_{\pm}^* $ and $z_{\pm}^*=\int dz/(c\mp v_0)^{-1}$,
for which the Polyakov approximation gives 
\be
\langle T_{\pm\pm}\rangle =-\hbar(12\pi)^{-1}C^{1/2}C^{-1/2}_{,\pm\pm}+\Delta_{\pm}\ . \ee
Here $C$ is the conformal factor for the 2D metric, $C=\rho_0 (c^2-v_0^2)c^{-3}$, 
and for the Unruh state $\Delta_{+}=0$, 
$\Delta_{-}= \hbar k^2/48\pi c^4$. 
\\Assuming as profile for the Laval  nozzle $A=\bar{A}+\beta z^2$, 
with $\bar{A},\beta $ constant, the classical solution reads
\bea
\label{classicsol}
\r_0&=&\bar{\r}e^{-\frac{v_0^2}{2c^2}}\ ,\, \bar{\r}=\mbox{const}\nonumber \\
z^2&=&\frac{\bar{A}}{\beta}\left[ 
\frac{c}{|v_0|}e^{\frac{v_0^2-c^2}{2c^2}} -1   \right]\ 
, \eea
where the sound velocity $c$ is taken to be constant. 
The location of the sonic horizon is $z=0$, where $v_0=-c$, 
and the region $z<0$ is the sonic black hole for which $|v_0|>c$.
These expressions should be regarded as describing the classical would be
asymptotic configuration of the fluid resulting from the (time dependent) 
formation of a sonic hole \cite{balisovi}. 
\\ Expanding the stress tensor and the background quantities near the horizon $z=0$ 
one eventually arrives \cite{lungo} to the following solution for the velocity
\be
\label{v}
v=v_0+ v_q \simeq -c+c\k z -\frac{c}{6}\k^2 z^2+\epsilon(b_1+c_1\k z)\k t
\ ,
\ee
where 
$\epsilon=\hbar /(\bar{A}^2\bar{\r}e^{-1/2}c)$ is the dimensionless expansion parameter and 
$b_1=9\gamma/2$,  
$c_1=-304\gamma/15$, where $\gamma=\bar{A}c^2\kappa^2/24\pi$,  and 
$\k=\sqrt{\beta/\bar{A}}$
has dimension $L^{-1}$ and is related to the surface gravity $\kappa=k/c^2$.\\
% For the density we find
% \be
% \label{rho}
% \r=\r_0+\r_1=\bar{\r} e^{-\half}
% [1+\k z +\epsilon (\alpha +\delta \k z)t]\ , 
% \ee
% with $\alpha>0$, $\delta <0$ \cite{lungo}.
% \be
% \alpha =\frac{\k^3 cA_0 }{5 \pi }\ , \ \ \ 
% \delta \simeq -3\alpha \ . 
% \ee
The solution is valid for $\k z \ll 1$ and $c\k t\ll 1$.\\
The boundary conditions imposed on the backreaction equations have been chosen 
so that at some given time (say $t=0$) the evaporation is switched on starting 
from the classical configuration $(v_0,\r_0)$, 
i.e. $\r_q(z,t=0)=v_q(z,t=0)=\f_q(z,t=0)=0$. 
\\ Inspection of eq.(\ref{v}) shows the net effect of the 
backreaction. Being 
% $b_1>|c_1|$ and
$\k z \ll 1$, from eq.(\ref{v}) 
we have $v_q>0$, i.e. the fluid is slowing down. 
This goes with an decrease of the density ($\r_q<0$) in the same limit \cite{lungo}.\\ 
Eq.(\ref{v}) allows also to follow the evolution of the acoustic horizon. 
Being the horizon defined by $v=-c$, this yields 
\be \label{zh}
z_H\simeq -\frac{\epsilon  b_1 t}{c}\ , 
\ee
i.e. the horizon is moving to the left with respect to the nozzle: 
the hypersonic region shrinks in size. The coefficient $b_1$ determining the quantum 
correction to the velocity (\ref{v}) and hence the evolution of the horizon is just the gradient
of the additional chemical potential related to the expectation value of the trace evaluated at $z=0$. This should be
compared to the black hole case where the evolution of the horizon is determined by the energy flux
($\dot M \propto \langle T^r_{\ t}\rangle$ in spherical symmetry \cite{York}).  
While in the latter case Hawking radiation occurs at the expense of the gravitational energy of 
the black
hole, in the fluid phonons emission takes away kinetic energy from the system.   \\ 
From eqs.(\ref{v}, \ref{zh}) one can evaluate 
the correction to the emission temperature  
\be
\label{T}
 T_U=\left. \frac{\hbar}{2\pi }\frac{\p v}{\p z}\right|_{z_H}
 =\frac{\hbar c}{2\pi }\k \left[ 1-\frac{563\epsilon}{720\pi}\kappa^3 c \bar{A} 
t\right]\ . 
\ee
Using particular values for liquid helium, the fractional change of the temperature per unit time is of order $10^{-8}\; s^{-1}$.\\ 
The expression we have obtained is rather significative. Unlike a Schwarzschild black hole the emission temperature for a sonic black hole decreases in time as the radiation emission proceeds.  This behaviour is reminiscent of the near-extremal Reissner-Nordstr\"om black hole ($M\stackrel{\sim}{>}|Q|$) where $Q$ is the conserved charge of the hole. As the mass decreases  because of Hawking evaporation, the Hawking temperature decreases as well, vanishing when $M=|Q|$. This ground state is approached in an infinite time (third law of black hole thermodynamics). One can conjecture a similar behaviour for the sonic hole with a vanishing temperature ground state. 
\\ The basic question which remains open is to what extent our results 
do depend on the dispersion relation assumed (free scalar field), which 
ignores short distance corrections due to the molecular structure of the fluid, 
and on the covariant regularization scheme used to subtract ultraviolet divergences. 
These two aspects are deeply connected. 
For the Polyakov theory we have used in the 2D backreaction equations (\ref{back1}, 
\ref{back2}), Jacobson \cite{jacobson} has argued that, within a covariant regularization scheme, 
no significant deviation from the usual expression for the trace anomaly and the components
of $\langle T_{ab}^{(2)}\rangle$ are expected if one introduces a cutoff at high frequencies.   
%Although 
%it has been shown that (under some hypothesis \cite{unruh04}) the spectrum of the emitted phonons is basically unaffected by the 
%dispersion relation \cite{jacobson}, no one knows what happens to 
%observables like $\langle T_{\mu\nu}\rangle$ which are in addition sensitive to the regularization scheme. 
%In a purely 2D context, Jacobson \cite{jacobson3} has argued that, within a covariant regularization scheme, no significant deviation from the usual expression for the trace anomaly and the flux are expected if one introduces a cutoff at high frequencies.
However, for the hydrodynamical system we have considered, covariance is a symmetry of the 
phonons low energy effective theory only, which is broken 
at short distance.  
Hence non covariant terms depending on the microscopic physics are expected to show up in the effective action
and are crucial for a correct description of the unperturbed quantum vacuum of the fluid. 
However the expectation values $\langle T_{\mu\nu}\rangle$ entering the backreaction equations 
(\ref{eqbackA}, \ref{eqbackB}) 
do not represent the energy momentum of the fluid quantum vacuum.
They describe instead the perturbation of the stationary vacuum (whose energy is strictly 
zero \cite{volovik}) induced by inhomogeneities and by the time dependent formation
of the sonic hole which triggers the phonons emission. In this paper we have assumed that these deviations can be computed 
within the low energy theory. This situation is not unusual. 
%In the traditional Casimir effect it is well known that the final finite result for the vacuum energy does not depend
%on the high energy frequency modes and is calculated directly from the low energy theory only.
%There are some examples for quantum liquids where this holds (see G. Volovik in Ref. \cite{libro}).    
Casimir effects are well known examples of vacuum disturbances caused by the presence of boundaries.
It happens that the Casimir energy is often (but not always, 
see G. Volovik in Ref. \cite{libro})
independent on the microscopic physics and can be calculated within the framework of the low
energy theory. This happens because, while low frequency modes are reflected by the 
boundaries, for the high energy
ones the wall is transparent.  They produce a divergent contribution to the vacuum energy which 
is canceled by a proper regularization scheme and does not affect the finite result. We have assumed
that a similar decoupling happens for the acoustic black hole.   
The check of our hypothesis would require an analysis of the quantum system within 
the microscopic theory which takes into account
the time dependent non homogeneous formation of the sonic hole. 
This is for the moment beyond computational capability.
Anyway, it has been shown that modifications of the dispersion relations, to take into account 
short-distance behaviour of the high-energy
modes, basically do not affect the spectrum of the emitted phonons \cite{jacobson2}. 
This is not a proof that observables like $\langle
T_{\mu\nu}\rangle$ are also unaffected by short-distance physics. However, it can be an illuminating 
 hint taking in mind what $\langle
T_{\mu\nu}\rangle$ does really represent and the indications coming from the Casimir effect. 
%Strictly speaking our results are consistent, within the limit of the 2D approximation, for systems where the functional integration over the fluctuation fields is dominated by configurations satisfying relativistic dispersion relation.\\
%The advantage of analogue models with respect to gravity is that a short distance theory is in principle known (for example $^3He$ interacting atoms). The major daunting problem is how to deal, in the microscopic theory, with the time-dependent non homogeneous motion of the fluid leading to the formation of an acoustic black hole and the consequent phonons emission. Modified dispersion relation for the effective theory are an interesting tool to encode the proper short distance behaviour. 
%Given the present limited understanding on how high energy modes  interact with the underlying medium, our work should be simply regarded  as a first attempt to tackle the backreaction problem for sonic black hole, an attempt performed using available technical tools developed in quantum field theory in curved space with all their intrinsic  limitations. It is obvious that more accurate analysis are needed and will hopefully be performed in the not too distant future. However the behaviour which emerges from our work, i.e. slowing down of the fluid  because of  energy dissipation by phonons emission and consequent decreasing of the emission temperature, is physically quite reasonable. It is rather unlikely that these features are just an artifact of the approximation we have used and therefore we expect them to be confirmed by future works.
\\ \noindent {\bf Acknowledgements}: We thank R. Emparan, L. Garay, T. Jacobson, M. Maio, R. Parentani 
and J. Russo for useful comments and E. Berti for help
with the figure.

%%%%%%%%%%%%%%%%%%%%%%%%%%%%%%%%%%%%%%%%%%%%%%%%%%%%%%%%%%%%%%%%%%%%%%%%%%%%

%\end{multicols}
\end{document}